%% file: main.tex
\documentclass[sigconf]{acmart}
\AtBeginDocument{%
  }
\setcopyright{none}
\renewcommand\footnotetextcopyrightpermission[1]{}

\input{config}

\title{SmartPoC: Generating Executable and Validated PoCs for Smart Contract Bug Reports}

\author{Longfei Chen}
\affiliation{%
  \institution{Tsinghua University}
  \city{Beijing}
  \country{China}
}

\author{Ruibin Yan}
\affiliation{%
  \institution{People's Public Security University of China}
  \city{Beijing}
  \country{China}
}

\author{Taiyu Wong}
\affiliation{%
  \institution{Tsinghua University}
  \city{Beijing}
  \country{China}
}

\author{Yiyang Chen}
\affiliation{%
  \institution{Tsinghua University}
  \city{Beijing}
  \country{China}
}

\author{Jialai Wang}
\affiliation{%
  \institution{National University of Singapore}
  \city{Singapore}
  \country{Singapore}
}

\author{Chao Zhang}
\authornote{Corresponding author.}
\affiliation{%
  \institution{Tsinghua University}
  \city{Beijing}
  \country{China}
}

\begin{document}

\input{0-abstract}

\maketitle

\input{1-introduction}

\input{2-background}

\input{3-methodology}
\input{4-implementation}

\input{5-evaluation}

\input{6-discussion}

\input{7-related-works}

\input{8-conclusion}

\input{9-data}

\bibliographystyle{ACM-Reference-Format}
\bibliography{ref}

\appendix


\end{document}

%% file: config.tex
\usepackage{graphicx} 
\usepackage{enumitem}
\usepackage{booktabs}
\usepackage{amsmath}
\usepackage{xspace}
\usepackage{xcolor}
\usepackage{tabularx}
\usepackage[ruled,vlined,linesnumbered]{algorithm2e}
\usepackage{makecell}
\usepackage{multirow}
\newcommand{\sysname}{{\tt SmartPoC}\xspace}
\newcommand{\sysnames}{{\tt SmartPoC}'s\xspace}
\newcommand{\enginename}{{GRE-Engine}\xspace}

\PassOptionsToPackage{hyphens}{url}\usepackage{hyperref}
\usepackage{url}
\usepackage{pgf}
\usepackage[caption=false,font=footnotesize]{subfig}

\usepackage{xcolor}       
\usepackage{amsmath} 

\usepackage{tikz}        
\usepackage[framemethod=tikz]{mdframed}  
\newmdenv[
  leftmargin=0pt,
  rightmargin=0pt,
  skipabove=6pt,
  skipbelow=6pt,
  backgroundcolor=black!7, 
  linecolor=black,
  linewidth=0.5pt,
  roundcorner=4pt,
  innerleftmargin=6pt,
  innerrightmargin=6pt,
  innertopmargin=4pt,
  innerbottommargin=4pt
]{rqanswer}
\newcommand{\rqbox}[2]{\begin{rqanswer}\textbf{Answer to RQ#1:} #2\end{rqanswer}}

%% file: 0-abstract.tex
\begin{abstract}

Smart contracts are commonly audited through static analysis to explore vulnerabilities. However, static approaches typically produce heterogeneous findings rather than reproducible, executable proof-of-concept (PoC) test cases, leading to costly and ad hoc manual validation.
Large language models (LLMs) offer a promising way to translate audit reports into PoC test cases, but face three major challenges: noisy inputs, lack of execution grounding, and missing runtime oracles. 
We present \sysname, an end-to-end approach for validating reported vulnerabilities in audit reports by generating and executing PoC test cases with automated exploitability verification.
\sysname first extracts a focused function-level slice from each report to reduce noise, centering on the key functions referenced in a finding and augmenting them with execution-relevant neighbors. To improve executability, we wrap LLM-based PoC synthesis in a generate–repair–execute loop, combining deterministic pre-execution sanitization with feedback-driven post-execution debugging. We further use differential verification as an oracle to confirm the exploitability of generated test cases. 
On the \textsc{SmartBugs-Vul} and \textsc{FORGE-Vul} benchmarks, \sysname achieves confirmation precision of 98.32\% and 98.65\%, with recall of 84.17\% and 85.28\%, respectively.
On a recent Etherscan verified-source corpus, \sysname confirms 64 bugs from 545 audit findings at an average cost of \$0.03.

\end{abstract}

%% file: 1-introduction.tex
\section{Introduction}

Smart contracts have become critical infrastructure for decentralized applications, but their immutability and direct control over funds make vulnerabilities uniquely costly~\cite{luu2016making,zhou2023sok}. Numerous tools perform static analysis~\cite{sun2024gptscan,yu2025smart,feist2019slither,weiss2024analyzing,mueller2018smashing} on contract code and generate vulnerability reports spanning a broad range of bug classes. However, static findings often exhibit high false-positive rates~\cite{ghaleb2022towards,sendner2024large,li2024static,lin2025large} and rarely include executable evidence. Consequently, practitioners have to conduct costly, ad hoc manual verification to determine whether reported issues are truly exploitable. We refer to this disconnect between detection and confirmable exploitability as the \emph{verification gap}.

In practice, auditors rely on executable proof-of-concept (PoC) tests to validate static findings. A PoC reproduces the vulnerable condition in a realistic runtime environment, encoding both (i) a triggering call sequence and (ii) an oracle that verifies the expected effects, making exploitability reproducible within testing frameworks~\cite{foundry2025,hardhat2025,verma2022application}. Although prior systems~\cite{jin2022exgen,krupp2018teether,zhang2020ethploit,rodler2023ef} have explored synthesizing exploit traces via fuzzing, symbolic execution, and other program analyses, they typically depend on expert-crafted predicates and bug-specific oracles. These oracles generalize poorly to the heterogeneous vulnerabilities prevalent in modern DeFi protocols~\cite{zhang2023demystifying}. 
As a result, even with increasingly capable static analyzers, auditors still author most PoCs manually. This process is labor-intensive and scales poorly with the growing volume of reported findings.

Large Language Models (LLMs) provide a promising avenue for this task.
With strong code understanding and synthesis capabilities~\cite{gervais2025aiagentsmartcontract,xiao2025prompt,wu2024advscanner,zhang2025automated,liu2024propertygpt,andersson2025poco}, they can, in principle, transform textual audit findings into executable PoC tests.
However, end-to-end, validated PoC generation from generic static-analysis outputs still faces three persistent challenges:
(i) \textbf{Noisy inputs to LLMs.} Vulnerable contracts often include substantial functionality unrelated to the reported issue, resulting in long and unfocused prompts. This long, noisy context prevents the model from concentrating on the true trigger conditions and consequently degrades PoC generation quality.
(ii) \textbf{Lack of execution grounding.} LLMs may generate unsupported APIs or incorrect environment assumptions, causing compilation failures. More critically, they may override target methods, inadvertently patching the bug or creating a fake one, so the “PoC” validates the LLM-written code rather than the original contract logic.
(iii) \textbf{Lack of a bug oracle.} There is no general, reliable mechanism to determine whether a PoC execution truly triggers the reported bug, which hinders scalable verification of audit findings.

In this work, we present \sysname, an end-to-end framework that validates smart contract vulnerability findings through PoC generation and execution.
The core idea is to build a verifiable system that harnesses LLMs for reliable PoC generation. To address the challenges above, \sysname consists of three key components:
(i) \textbf{Bug-context extraction (BCE).} To reduce input noise, BCE extracts a focused function-level slice centered on the key functions referenced in a finding, and augments it with semantic and structural links. The resulting context captures the execution prerequisites for PoC generation while keeping prompts compact.
(ii) \textbf{Generate-repair-execute engine (GRE-Engine).} To improve executability, GRE-Engine performs iterative synthesis with two repair modules. A pre-execution sanitizer normalizes the environment and prevents unintended contract rewrites, reducing the risk that generated PoCs mask the original bug. A post-execution debugger collects compiler and runtime diagnostics and feeds them back to the generator, enabling targeted repairs until the PoC executes successfully.
(iii) \textbf{Action-state differential verification (DV).}  
DV uses the GRE-Engine to instrument each PoC with pre- and post-trigger state queries, executes the test, and compares the resulting snapshots. A differential state change consistent with the vulnerability semantics indicates successful exploitation, providing an automatic oracle suitable for large-scale PoC verification.

We instantiate \sysname with DeepSeek-R1~\cite{guo2025deepseek} as the default PoC generator and evaluate it on three datasets: the traditional benchmark \textsc{SmartBugs-Vul}, the expert audit-report corpus \textsc{FORGE-Vul}, and the real-world \textsc{Latest-114} from Etherscan verified-source contracts. On the ground-truth benchmarks, \sysnames achieves a precision of 98.32\% and 98.65\%, with a recall of 84.17\% on \textsc{SmartBugs-Vul} and 85.28\% on \textsc{FORGE-Vul}, respectively. 
\sysname validates 507 and 282 findings produced by six static analyzers~\cite{sun2024gptscan,wei2024ftsmartaudit,yu2025smart,feist2019slither,weiss2024analyzing,mueller2018smashing} on the two benchmarks, with manual adjudication yielding PPV/NPV of 76.47\%/70.39\% and 80.00\%/89.72\%, respectively.
In the real-world study, \sysname validates 236 of 545 static-tool findings on \textsc{Latest-114}, achieving an audited precision of 75.29\% and an average validation cost of \$0.03 per finding. We further conduct ablation studies to quantify the contribution of each component and assess robustness to the LLM backend by substituting GPT-5-mini~\cite{achiam2023gpt} for DeepSeek-R1. Overall, the results show that \sysname effectively translates static-analysis findings into executable, runtime-grounded validation.

We summarize our main contributions as follows:
\begin{itemize}
    \item We propose an end-to-end approach that dynamically validates generic static-analysis reports by generating and executing PoCs.
    \item We design and implement \sysname, which integrates three key components—BCE, GRE-Engine, and DV—to robustly handle noisy inputs and lack of execution grounding, and to provide automated runtime oracles.
    \item We integrate \sysname with multiple static analyzers and validate 64 real bugs in the latest Ethereum corpus at an average cost of \$0.03 per validated finding.
    \item We release a large-scale dataset of 382 triplets linking contract source code, natural-language bug descriptions, and execution-validated PoCs. To the best of our knowledge, this is among the largest publicly available datasets that combine these three artifacts.
\end{itemize}

%% file: 2-background.tex
\section{Background}\label{sec:background}

\subsection{The Verification Gap of Static Analysis} 
Many smart-contract vulnerability detection solutions rely on static analysis, including rule- and pattern-based analyzers~\cite{weiss2024analyzing,feist2019slither,mossberg2019manticore,tsankov2018securify,schneidewind2020ethor,mueller2018smashing,luu2016making} as well as learning- or LLM-assisted approaches~\cite{yu2025smart,ma2025combining,sun2024gptscan,wei2024ftsmartaudit,sun2024llm4vuln,chen2025smarttrans}. Despite their differences, these tools share a common limitation: they operate on static code snapshots and typically output descriptive findings,
such as warnings, locations, and brief explanations, rather than executable evidence that can be validated at runtime.
This limitation reflects the inherent difficulty of recovering dynamic dependencies through static reasoning alone.
For example, a detector may flag a “reentrancy-like” pattern in isolation, yet the vulnerability may only materialize under a specific inter-contract initialization and liquidity sequence, which a static report cannot faithfully capture. In addition, static analyzers such as Slither~\cite{feist2019slither} include many detectors spanning diverse bug classes and scan for extensive sets of patterns, often producing large numbers of candidate findings with heterogeneous descriptions. As a result, they can exhibit high false-positive rates~\cite{ghaleb2022towards,sendner2024large,li2024static,lin2025large} and impose substantial manual effort for triage and confirmation. We refer to this growing divide between static detection and confirmed exploitability as the \emph{verification gap}.

\begin{figure}[!tbp]
        \centering
        \includegraphics[width=\linewidth]{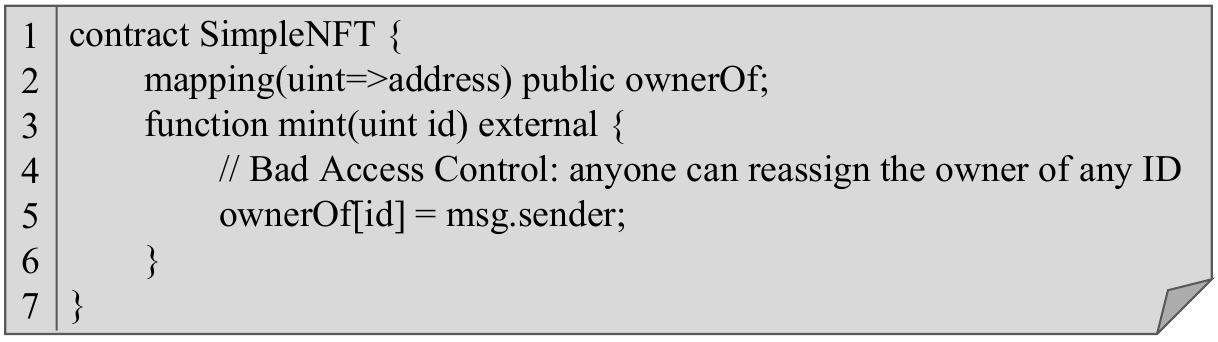}
        \caption{Access-control lapse requiring semantics-aware validation. In \texttt{mint}, any caller can reassign the owner of any token ID; confirming exploitability needs an ownership oracle (e.g., \texttt{ownerOf[id] == attacker}) rather than generic profit signals.}
        \label{fig:background-example}
\end{figure}

\subsection{The Format of Static Reports}
In practice, static vulnerability reports primarily originate from two sources. The first is audit reports written by human experts, which are often semantically rich and may include the vulnerable location, a textual explanation of the issue, potential consequences, and mitigation recommendations. The second is tool-generated findings, which are typically more structured and concise, often containing only a vulnerability label and a code location. Recent LLM-based auditing tools further blur this distinction by enriching structured findings with natural-language descriptions.
Despite these differences, we abstract all such inputs as generic findings. 

For \sysname, each generic finding provides two essential signals: the suspected location and a textual description of the vulnerability type or semantics. The location may be specified as a line range, a function name, or a program point that can be mapped to the corresponding function, whereas the description may range from a short label to a more detailed natural-language explanation.

\begin{figure*}[!t]
        \centering
        \includegraphics[width=1\linewidth]{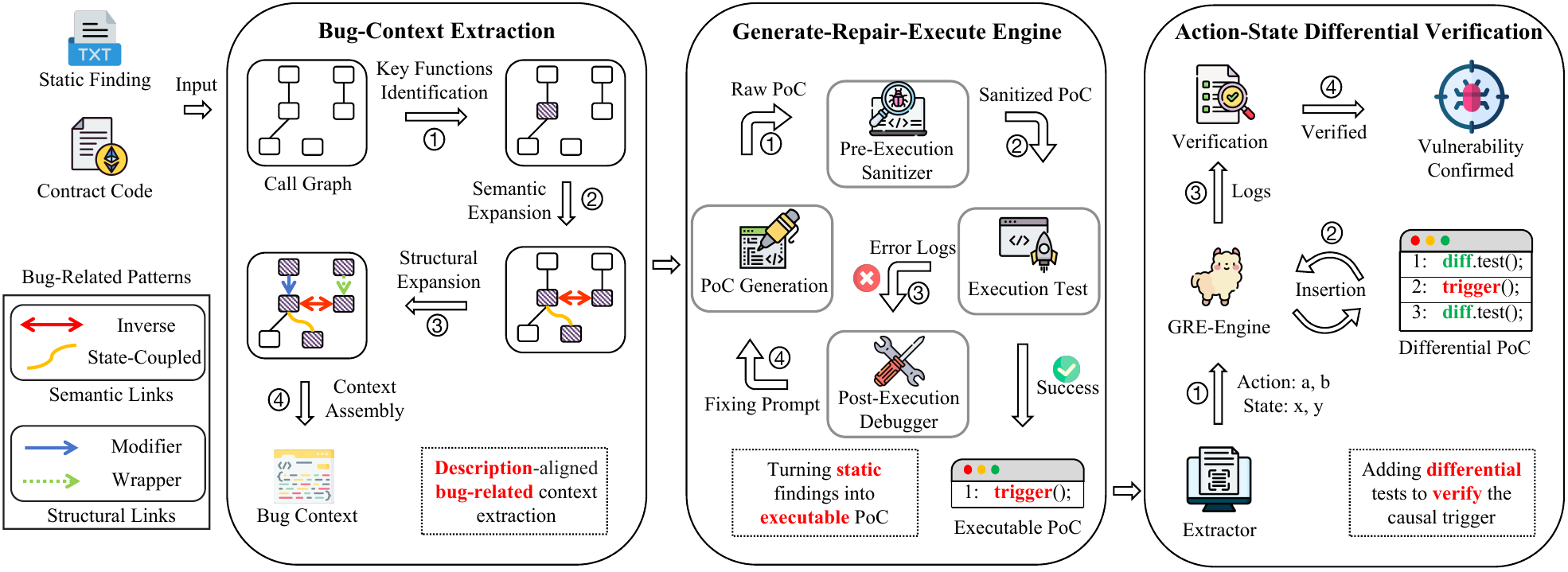}
        \caption{The high-level workflow of \sysname.}
        \label{fig:workflow}
\end{figure*}

\subsection{Proof-of-Concept and Runtime Oracle}

\textbf{PoC definition.}
A PoC is an executable artifact that demonstrates whether a reported vulnerability can be triggered in a realistic environment. Unlike descriptive static warnings, a PoC makes exploitability testable by encoding (i) a triggering call sequence and (ii) an oracle that validates the expected effects, thereby turning descriptive findings into execution-grounded evidence. In practice, PoCs are commonly instantiated either as vulnerable transaction sequences (i.e., multi-transaction execution traces) or as reproducible test harnesses implemented in standard testing frameworks such as Foundry~\cite{foundry2025}, Hardhat~\cite{hardhat2025}, and Truffle~\cite{verma2022application}.

\textbf{Runtime oracle.}
Executing a trigger alone demonstrates only path feasibility, not exploitability. Confirming a vulnerability requires an oracle to determine whether the observed outcome constitutes a security-relevant misuse. Profit-based criteria are natural for profit-driven DeFi exploits, but they do not generalize to semantic violations such as privilege escalation, unauthorized state transitions, or broken invariants.
Figure~\ref{fig:background-example} shows a typical access-control flaw: the \texttt{mint} function allows an arbitrary caller to assign ownership of any token ID. Successful exploitation in this case is not necessarily immediately profitable; instead, it requires a semantics-aware identity or uniqueness oracle. For example, one may assert \texttt{ownerOf[id] == attacker} and optionally \texttt{preOwner != attacker} to detect the bug. 
Empirically, approximately 80\% of real-world smart contract vulnerabilities are logic-centric~\cite{zhang2023demystifying}. For such issues, profit-based oracles alone are often insufficient; instead, validation requires semantics-aware runtime checks over security-relevant state changes, privilege violations, or broken invariants. In practice, these checks remain largely handcrafted, making exploitability validation both time-consuming and error-prone.

%% file: 3-methodology.tex
\section{Methodology}

\begin{figure}[!tbp]
        \centering
        \includegraphics[width=1\linewidth]{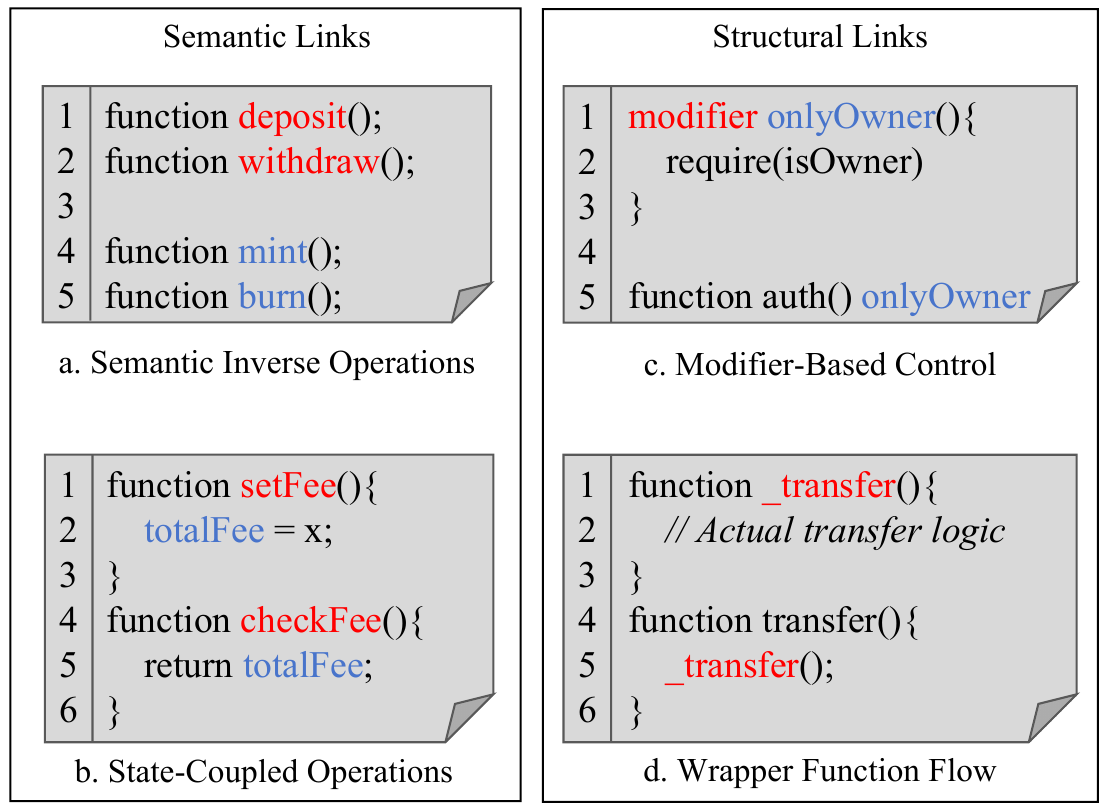}
        \caption{Bug-related patterns in smart contract code.}
        \label{fig:bce-example}
\end{figure}

\subsection{Overview}

The workflow of \sysname is shown in Figure~\ref{fig:workflow}. \sysname transforms the inputs into executable, validated PoCs in three stages:
(1) \textbf{Bug-Context Extraction (BCE)} uses the static finding to locate the report-referenced functions. It then expands these functions via semantic and structural links and assembles the corresponding function-level slice as the LLM context. 
This context preserves bug-relevant semantics and common setup prerequisites, while avoiding the noise of the full project. 
(2) \textbf{Generate-Repair-Execute Engine (GRE-Engine)} converts the finding and BCE context into an executable PoC by wrapping LLM synthesis in an iterative generate--repair--execute loop.
It includes two repair modules.
A pre-execution sanitizer normalizes project-specific dependency details and blocks test-level non-faithful edits that could override the target behavior.
A post-execution debugger leverages compiler and runtime diagnostics to guide regeneration until the PoC executes or the retry budget is exhausted. 
(3) \textbf{Action--State Differential Verification (DV)} provides execution-grounded evidence of exploitability through controlled comparisons.
The core idea is to bracket the PoC trigger with a concrete action (i.e., a call or short call sequence) and monitor a bug-related state (i.e., public variables or view functions).
DV confirms a PoC only when the observed pre/post difference is explicit and consistent with the report.
For each finding, \sysname returns the PoC together with the differential evidence.
If validation fails, \sysname marks the finding as not validated under our oracle.

\subsection{Bug-Context Extraction} \label{sec:bug-context-extraction}
BCE aims to construct a compact context that captures the information required for PoC execution.
It reduces prompt noise by extracting a focused function-level slice instead of feeding the entire contract to the LLM. It starts from the functions referenced by the finding and includes a one-hop expansion on the call graph. 
However, bug-local context alone is often insufficient as a working PoC must first drive the contract into an exploitable pre-state. To support such runtime prerequisites, BCE further expands the function set using empirically observed patterns.

We summarize four common links between functions that frequently arise in audited contracts, as illustrated in Figure~\ref{fig:bce-example}.   
\emph{(a) Semantic inverse operations.} Some contracts expose complementary operations with opposite effects. A PoC may require the inverse operation to establish the pre-state before triggering the reported entry point. For example, a report may highlight \texttt{withdraw}, but the PoC must call \texttt{deposit} first to fund the attacker.
\emph{(b) State-coupled operations.} Some operations interact through shared state, while updates and validations are implemented in different functions. A PoC may need both functions to make a state change observable and to reach the target path. For example, \texttt{setFee} updates a fee parameter, while the fee bound may be enforced by \texttt{checkFee} during use.
The next two are structural links that affect how the PoC should invoke the contract.
\emph{(c) Modifier-based control.} Modifiers commonly encode authorization and validity checks that gate execution~\cite{fang2023beyond}. A PoC should account for these checks to avoid failing before reaching the vulnerability trigger. For example, \texttt{onlyOwner} requires the PoC to impersonate the owner or to set ownership accordingly.
\emph{(d) Wrapper function flow.} A common Solidity design pattern is to expose a public-facing API that forwards to underscore-prefixed internal helpers implementing the core logic~\cite{solidity2025}. A PoC should preserve this call structure to interact with the contract as intended. For example, \texttt{transfer} may delegate to \texttt{\_transfer} where balance updates occur.
Together, these patterns help retain execution-critical context for exploit reproduction, which are not reliably captured when slicing only around local control-flow. To estimate how often these links appear in practice, we perform a lightweight lexical scan over 373 real-world audit projects from FORGE~\cite{chen2025forgellmdrivenframeworklargescale}. 
We find that 97.61\% of projects contain at least one of the four links, indicating that these patterns are prevalent in real codebases.

\begin{figure}[!t]
        \centering
        \includegraphics[width=0.87\linewidth]{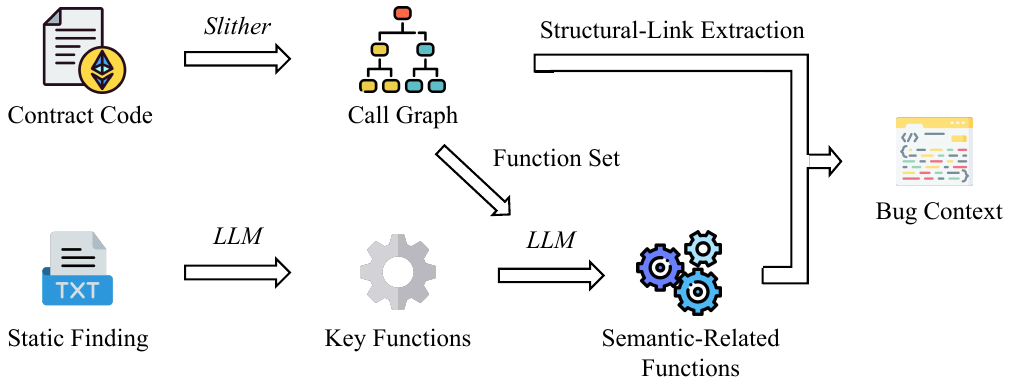}
        \caption{Workflow of Bug-Context Extraction.}
        \label{fig:bce-workflow}
\end{figure}

As shown in Figure~\ref{fig:bce-workflow}, BCE is seeded with the contract code and a corresponding static finding, where each finding is represented by two fields: a location and a vulnerability description. For normalized findings produced by static analyzers, BCE directly maps the reported location to the corresponding function(s). For more heterogeneous findings, such as audit reports or LLM-generated findings, BCE identifies the key function(s) from the vulnerability description.
We first run \texttt{Slither}~\cite{feist2019slither} to recover the call graph and a canonical function list of the project.
Next, instead of directly matching function names mentioned in the report, we ask the LLM to identify the referenced functions, which we denote as \emph{key functions}.
Reports often mention functions in non-canonical forms---e.g., case variations, minor typos, or formatting differences---which makes string matching brittle.
We therefore formulate key-function identification as a closed-set selection task: we provide the LLM with the recovered function list and require it to select the report-referenced functions only from this list.
We then verify the selections against the same list and drop any out-of-set symbols, further mitigating hallucinated function names. 
To expand semantic links, we use few-shot~\cite{brown2020language} prompts with a set of pre-defined \emph{semantic inverse} and \emph{state-coupled operations} drawn from common contract idioms. 
Guided by these patterns, the LLM selects semantically related functions from the recovered function list.
We include the selected functions as additional execution context so that GRE-Engine can access prerequisite states beyond the bug-referenced entry points.

After that, we expand the context along control-flow structure rooted at these functions. We first include the call-graph edges among the selected functions, and then add their direct callees and callers. To handle wrapper patterns, we recursively include underscore-prefixed internal functions reachable from these callers and callees until we reach the non-underscored implementations. We also attach all referenced \texttt{modifier}s to preserve pre-/post-condition checks and access-control boundaries. 
To support correct initialization during generation, we include constructors and, when present, initializer routines for the contracts that define the sliced functions. We also extract from the code the project and build metadata needed for compilation, such as the compiler version and remappings. Finally, we deduplicate the resulting function set while preserving reachability order, and package everything into a compact bug-context bundle for subsequent PoC generation.

\begin{figure}[!t]
  \centering
  \includegraphics[width=0.9\linewidth]{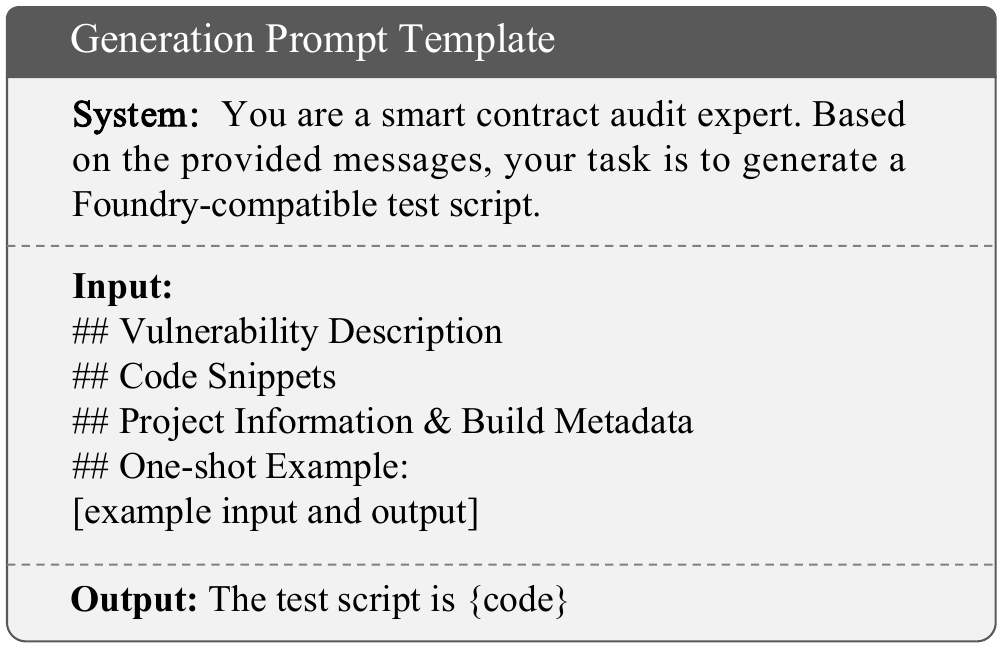}
  \caption{Prompt for PoC generation.}
  \label{fig:generation-prompt}
\end{figure}

\subsection{Generate-Repair-Execute Engine}\label{sec:poc-generation} 

\enginename turns textual vulnerability descriptions into executable PoCs via an automated generate–repair–execute loop that iteratively prompts an LLM. Its core consists of two repair stages: a pre-execution sanitizer and a post-execution debugger. The sanitizer applies deterministic rules to fix runtime dependencies and block non-faithful behaviors, while the debugger collects compiler and runtime failures and uses the resulting diagnostics to guide regeneration.

In practice, we observe three recurring failure modes when using LLMs to generate PoCs.
First, LLMs are often not framework-version aware. As libraries evolve, they may call deprecated APIs or use outdated import paths, leading to compilation failures~\cite{wang2024llms}.
Such version mismatches are difficult to resolve through prompting alone. 
Second, LLMs may ``cheat'' to satisfy the report narrative. They may override target methods inside the PoC and inject a vulnerable behavior to force the expected state change. In this case, the test exercises the overridden implementation rather than the original contract, producing a false confirmation even when the target code is not actually vulnerable. 
Third, a PoC may compile but still fail at execution time. A PoC must interact with the target project in a concrete runtime environment, and mismatched calls, missing setup steps, or incorrect assumptions can cause interaction failures. Resolving such failures typically requires dynamic feedback from compilation and test execution.

\begin{figure}[!t]
  \centering
  \includegraphics[width=0.9\linewidth]{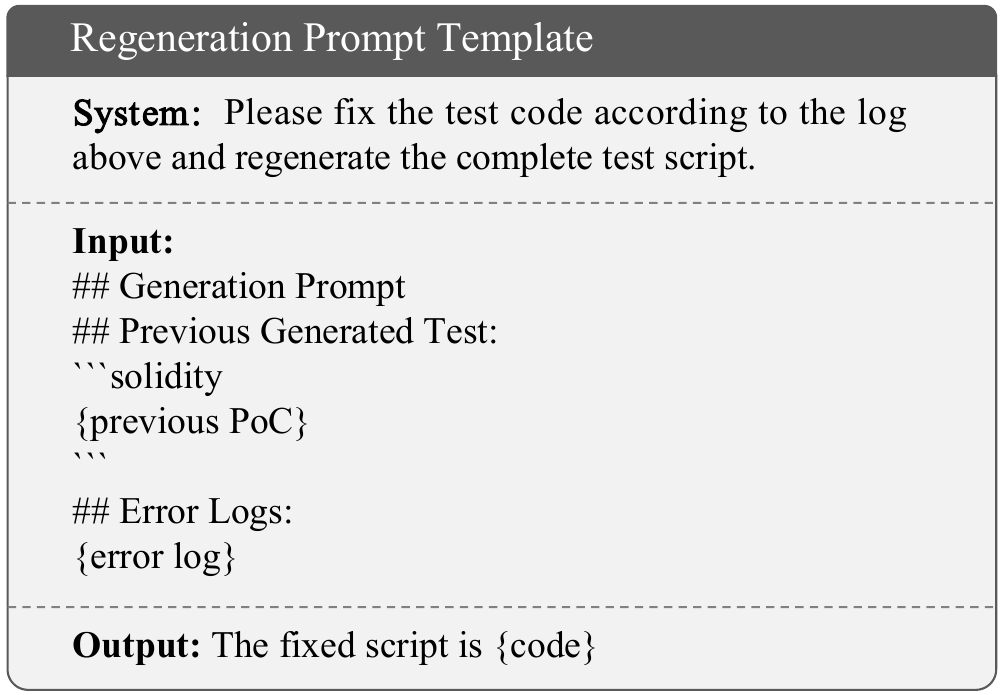}
  \caption{Prompt for post-execution debugger.}
  \label{fig:regeneration-prompt}
\end{figure}

To address these challenges, we organize PoC generation into an iterative loop. The process begins by prompting the LLM to generate an initial draft (Figure~\ref{fig:generation-prompt}). The draft then passes through a pre-execution sanitizer to address the first two issues. We select Foundry~\cite{foundry2025} as the testing framework and pre-adapt the environment across Solidity compiler versions. Specifically, we extract the compiler version and library identifiers from the draft's header directives and apply deterministic edits to ensure compatibility with the appropriate syntax and APIs. We also align the PoC with the target codebase by matching the target project's canonical function list against the PoC, removing conflicting or redundant re-definitions, and rewriting imports to correctly reference the required headers. These deterministic fixes substantially reduce compilation failures and enforce that PoCs interact with the original contract code, mitigating non-faithful transformations.
After the PoC passes pre-execution sanitizer, we run it in the test environment and record any compilation or runtime errors. In the post-execution debugger, we structure these diagnostics and prompt the LLM to revise the PoC accordingly (Figure~\ref{fig:regeneration-prompt}). We retain the original conversational context to anchor the model to the finding and intended exploit scenario. To keep each repair focused, we provide only the latest PoC draft and the diagnostics from the current round, omitting older debugging history. We repeat this loop until the PoC executes, or until the attempt budget is exhausted.

\subsection{Action--State Differential Verification}\label{sec:differential-verification}

While \enginename produces PoCs that compile, execute, and interact correctly with the target contract, action--state differential verification validates their execution outcomes and semantic correctness. We define a \emph{state} as a bug-relevant observable value (e.g., a public variable or view function result) obtained from execution, and an \emph{action} as the concrete interaction (e.g., a call or short call sequence) used to obtain that value.
By comparing the state before and after under matched controls, we determine whether the vulnerability produces a real, report-consistent effect. This approach leverages the LLM to tailor bug-specific oracles to complex protocol logic while keeping the final decision rule simple and auditable. It avoids prior work’s reliance on hand-crafted oracles and enables verification for generic static-analysis findings.

\begin{figure}[!t]
        \centering
        \includegraphics[width=0.9\linewidth]{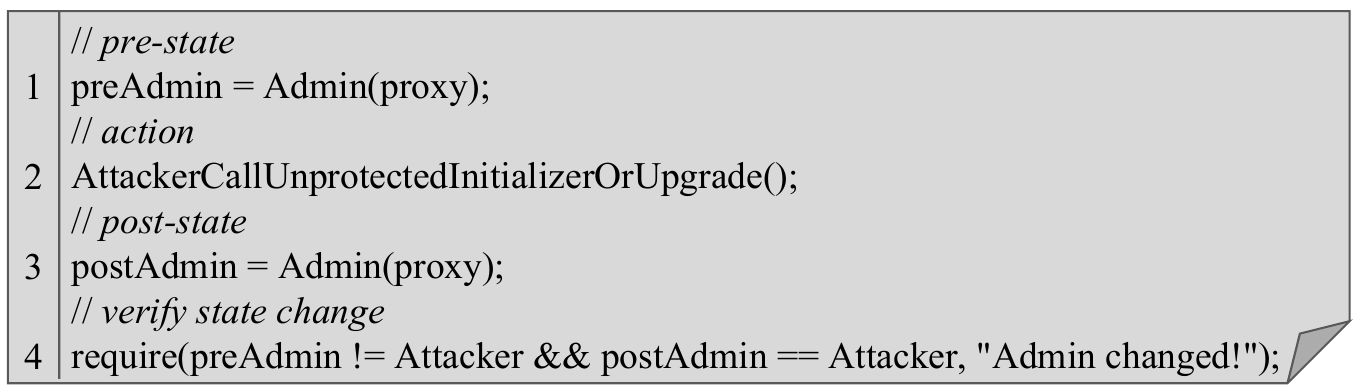}
        \caption{DV for proxy-control takeover. DV snapshots the proxy's upgrade authority (\texttt{admin}) before and after the initializer/upgrade trigger (Lines~1--3), and then checks whether control has been illicitly transferred to the attacker (Line~4).}
        \label{fig:differential-example}
\end{figure}

Although triggers and root causes vary across bugs, confirming exploitability requires execution-grounded evidence.
We therefore base DV on observable runtime traces, including queried state values and call outcomes, and compare them under matched controls.
Typical evidence includes (i) direct changes in variables (e.g., balances, ownership, and liquidation thresholds), and (ii) changes in function return values or effects (e.g., success flags of authorization checks). If we can localize such a state and test its pre-/post- values around the trigger, DV can often reduce validation to a single outcome-based check. Figure~\ref{fig:differential-example} illustrates a proxy-control takeover example. The trigger occurs on Line~2, where the attacker exploits an unprotected initializer or upgrade path to seize the proxy’s upgrade authority. We snapshot the proxy admin both before and after the trigger (Lines~1 and~3) and compare the two values on Line~4 to determine whether control has been illegitimately transferred to the attacker. In this setup, the attacker’s trigger sequence is the \emph{action}, and the proxy’s admin address is the observable \emph{state}.

\begin{figure}[!t]
  \centering
  \includegraphics[width=0.9\linewidth]{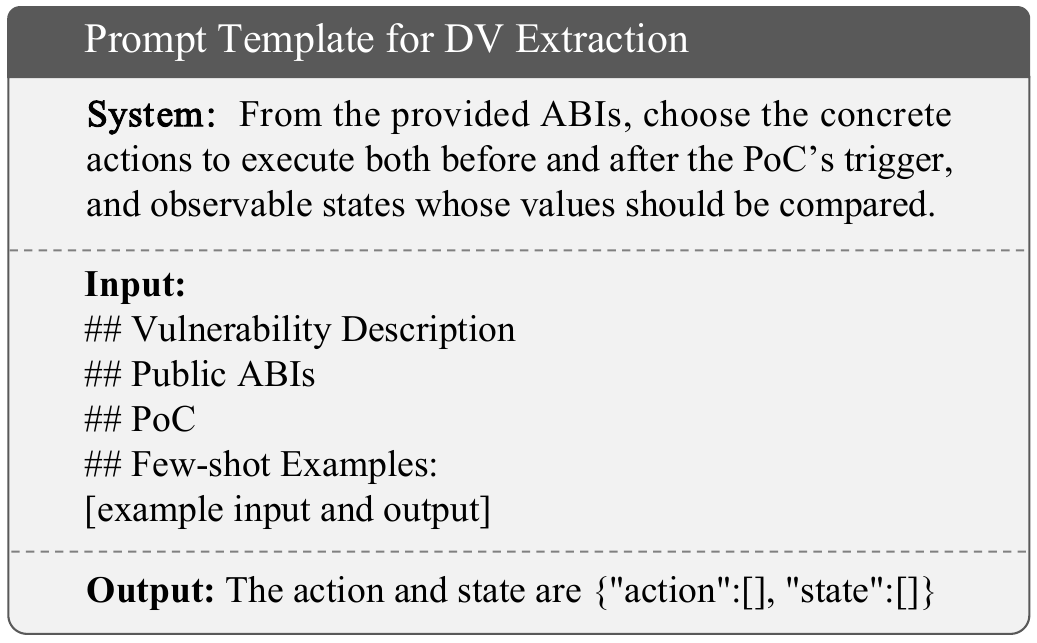}
  \caption{Prompt for DV extraction.}
  \label{fig:prompt-dv-extractor}
\end{figure}

To operationalize this idea, we design a differential verification pipeline that decides exploitability by checking localized, publicly observable state changes around the trigger. We follow three steps:

\noindent
(1) \textbf{Extraction.}
Given the finding description, the generated PoC, and the public ABIs (all externally visible functions and getters for public state variables) derived from the source code in Section~\ref{sec:bug-context-extraction}, we prompt the LLM to identify (a) an \emph{action} (one or a small sequence of function calls) that exercises the trigger and (b) the observable \emph{state} to monitor.
As shown in Figure~\ref{fig:prompt-dv-extractor}, the LLM selects candidate functions from the public ABIs for both the action and the state query.
This design leverages the LLM’s understanding of the vulnerability while constraining it to a contract-specific search space, thereby narrowing the validation oracle.

\begin{figure}[!t]
  \centering
  \includegraphics[width=0.9\linewidth]{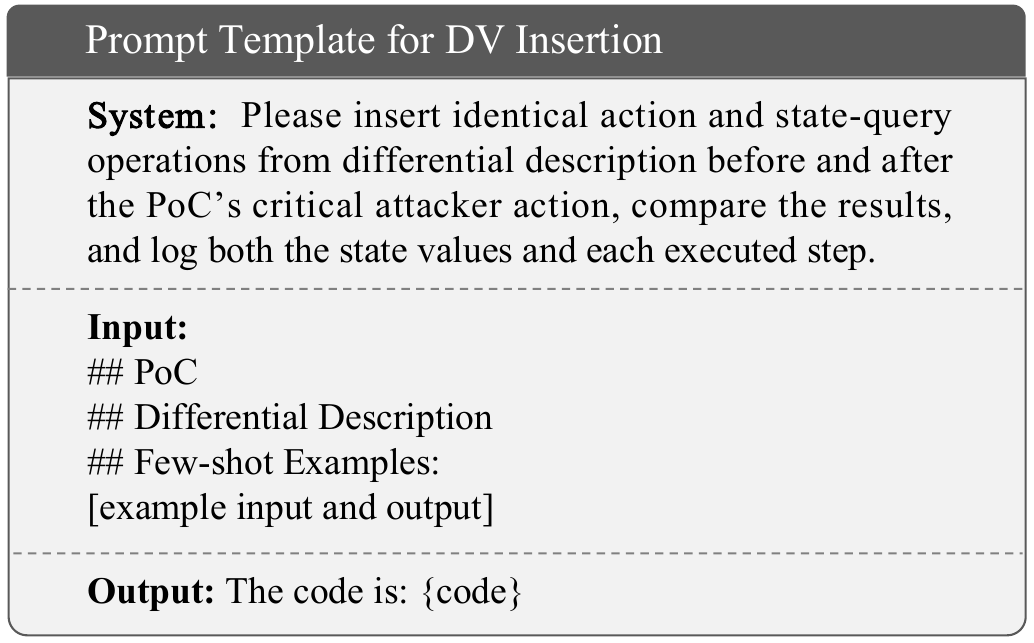}
  \caption{Prompt for DV insertion.}
  \label{fig:prompt-dv-insertion}
\end{figure}

\noindent
(2) \textbf{Insertion.}
We use GRE-Engine to materialize the selected action and state as code, producing an executable PoC with embedded differential checks.
As shown in Figure~\ref{fig:prompt-dv-insertion}, we provide the LLM with the original PoC and the extracted action/state pair and ask it to insert the corresponding calls and assertions into the test.
For each pre-/post-action segment and at the trigger, we instruct the LLM to insert structured log statements that record the executed step and the current state value.
In this way, the abstract oracle is instantiated as a concrete, executable test that can be run and inspected. During both extraction and insertion, we provide few-shot examples to help the LLM follow instructions and translate vulnerability semantics into causal decisions and concrete execution logic.

\noindent
(3) \textbf{Verification.}
After execution, we parse the structured logs to extract the pre- and post-state and compute the corresponding deltas. When a state change is observed, we further query an LLM to sanity-check whether the change constitutes evidence for the reported vulnerability. We label a PoC as successful only if it induces a vulnerability-relevant state change under this check. In this design, successful execution ensures that the exercised path is feasible, while the LLM-based sanity check validates the vulnerability semantics. Together, these steps automatically derive runtime oracles that validate LLM-generated PoCs and confirm exploitability.

%% file: 4-implementation.tex
\section{Implementation and Setup}

\subsection{Dataset}
Our evaluation uses three datasets: a community benchmark SmartBugs~\cite{ferreira2020smartbugs}, a corpus of projects from expert audit reports  FORGE~\cite{chen2025forgellmdrivenframeworklargescale}, and a real-world collection of Ethereum contracts Latest-114.

\noindent \textbf{SmartBugs} covers ten canonical vulnerability types, together with corresponding contracts and labels. Because Foundry provides limited support for older Solidity versions, we upgrade all contracts to Solidity $\geq 0.8$ through LLM-assisted, human-validated edits. During this process, we exclude vulnerability classes that no longer manifest after the upgrade (\emph{short address} and \emph{others}). For each retained sample, we use the dataset-provided description and map the location to the relevant key functions. The resulting benchmark contains 139 vulnerability samples, denoted as \textsc{SmartBugs-Vul}.

\noindent \textbf{FORGE} consists of project codebases paired with commercial audit reports. For each project, we retain the source code and report, and then: (i) extract only high- and critical-severity findings; (ii) perform batch compilation to exclude projects with incomplete or irreproducible code; (iii) segment each report into per-finding units with LLM assistance. We discard items that lack a concrete location. This process yields 428 vulnerability instances, denoted as \textsc{FORGE-Vul}, covering 24 CWE types.

To simulate false positives in static-analysis reports, we construct \textsc{SmartBugs-Fix} and \textsc{FORGE-Fix} by patching the underlying vulnerabilities while keeping the original textual descriptions unchanged. These fixed sets are used to evaluate whether \sysname\ can avoid confirming non-vulnerable code despite persuasive vulnerability narratives. For \textsc{SmartBugs-Fix}, we repair all vulnerable instances. For \textsc{FORGE-Fix}, because manual patching is costly, we patch a fixed budget of 100 instances. Specifically, we adopt CWE-stratified sampling. To ensure coverage of long-tail categories, we first select at least two instances from each CWE type. We then allocate the remaining budget proportionally according to the frequency of each CWE type and randomly sample instances within each stratum. All patches are produced independently by two domain experts and cross-validated. During patching, we preserve the intended contract functionality and keep function signatures unchanged to remain consistent with the original vulnerability reports.

\noindent \textbf{Latest-114} is collected from Etherscan's verified-source corpus through the public API~\cite{etherscan_verified_contracts}. On October 22, 2025, we queried the 150 most recently verified projects and reconstructed the corresponding build environments. After normalization, 114 projects compiled successfully, forming the \textsc{Latest-114} dataset. This dataset evaluates \sysname\ in a recent real-world setting by assessing how effectively it validates static-analysis findings on organically developed contracts deployed on-chain.

\subsection{Baselines}
Three recent studies are related to automated exploit generation for smart contracts, namely \textsc{A1}~\cite{gervais2025aiagentsmartcontract}, \textsc{PoCo}~\cite{andersson2025poco}, and \textsc{REX}~\cite{xiao2025prompt}. \textsc{REX} follows an iterative prompt-driven exploit-synthesis paradigm, \textsc{PoCo} adopts an agentic report-to-PoC generation paradigm, and \textsc{A1} targets an agentic exploit-discovery setting over deployed contracts. Since these works do not provide publicly runnable implementations, we construct controlled baselines that capture the representative PoC-generation workflows described in these studies.
\emph{Iterative} is an iterative baseline inspired by \textsc{REX}. It generates a PoC from the target contract and vulnerability report, executes the result, and iteratively revises it using compilation and execution feedback.
\emph{Agentic} is the baseline inspired primarily by \textsc{PoCo} and \textsc{A1}. It allows the LLM to plan, implement, execute, and debug the PoC itself.
We also include four simplified variants of \sysname: \emph{w/o BCE}, which uses the flattened full contract as context; \emph{w/o Semantic Links}, which retains only the control-flow-reachable slice from the key functions; \emph{w/o Pre-Repair}, which removes the pre-execution sanitizer; and \emph{w/o Post-Repair}, which removes the post-execution debugging module.

\subsection{Experimental Setup} \label{sec:setup}

\textbf{Framework settings.}
All experiments use the open-source reasoning model \emph{DeepSeek-R1-0528}~\cite{guo2025deepseek}. To assess robustness, we also run a subset of experiments with \emph{GPT-5-mini}~\cite{achiam2023gpt}. Unless otherwise noted, we use default decoding with temperature $0.3$ and set the \enginename\ retry budget to $B{=}5$. For PoC generation and execution, we use Foundry as the testing framework, relying on the latest release available during our experiments (\texttt{forge} \texttt{1.4.2-stable}).

\noindent \textbf{Static analysis tools.}
We select (i) three widely used rule-/pattern-based analyzers~\cite{feist2019slither,weiss2024analyzing,mueller2018smashing}, and (ii) three state-of-the-art LLM-assisted analyzers~\cite{sun2024gptscan,wei2024ftsmartaudit,yu2025smart}. We run each tool with default configurations with a 1800-second time limit, and retain only \emph{high-severity} findings. For each finding, we provide \sysname\ with the vulnerability narrative and the implicated functions. 
\emph{Slither}~\cite{feist2019slither} translates Solidity into SlithIR and performs rule-based checks augmented with data-/control-flow analysis (6k GitHub stars).  
\emph{Mythril}~\cite{mueller2018smashing} symbolically executes EVM bytecode to explore feasible paths and flag suspicious behaviors (4.1k Github stars).  
\emph{CPG Contract Checker (CCC)}~\cite{weiss2024analyzing} converts source into a Code Property Graph and uses declarative graph queries that encode ``vulnerability pattern + data/control-flow conditions + exceptions/mitigations,'' enabling precise structural matching.
\emph{GPTScan}~\cite{sun2024gptscan} defines per-class scenarios and properties for 10 logic-bug classes and prompts an LLM for classification and explanation, aided by lightweight static slicing and simple validation.  
\emph{FTSmartAudit (FT)}~\cite{wei2024ftsmartaudit} distills high-quality audit data and performs lightweight fine-tuning with a rationale-aware dual loss; an iterative filter–refine loop compresses a general LLM into a compact, audit-specialized model.  
\emph{SMART-LLAMA-DPO (DPO)}~\cite{yu2025smart} jointly learns vulnerability classification and explanation within a single framework, then applies direct preference optimization to align explanation quality with human.

%% file: 5-evaluation.tex
\section{Evaluation}
We aim to answer the following questions:
\begin{itemize}
\item \textbf{RQ1: Can \sysname correctly validate vulnerabilities reported in audit reports?}
Given benchmark and expert audit reports, we evaluate whether \sysname can correctly validate these reported issues.

\item \textbf{RQ2: Can \sysname serve as an execution-grounded verifier for analyzer findings?}
We run six analyzers to produce candidate findings on three datasets and then use \sysname to validate them. We further manually audit random samples of \sysname-confirmed and \sysname-rejected cases to estimate PPV and NPV.

\item \textbf{RQ3: How does \sysname perform in PoC generation compared with baselines?}
We compare \sysname with several representative PoC generation settings and evaluate the effectiveness of its key designs.

\end{itemize}

\subsection{Evaluation Metrics}
We evaluate \sysname using standard confusion-matrix counts---$\mathrm{TP}$, $\mathrm{TN}$, $\mathrm{FP}$, and $\mathrm{FN}$---computed by comparing \sysname's validation outcomes against expert adjudication. We treat findings that \sysname \emph{confirms} as vulnerable as \emph{positives}, and those it \emph{rejects} as \emph{negatives}. A confirmed case is counted as $\mathrm{TP}$ if the vulnerability is indeed exploitable; otherwise it is $\mathrm{FP}$. A rejected case is counted as $\mathrm{TN}$ if the code is non-vulnerable; otherwise it is $\mathrm{FN}$.
We report recall and predictive values:
$\mathrm{Recall}=\frac{\mathrm{TP}}{\mathrm{TP}+\mathrm{FN}}$,
$\mathrm{PPV}=\frac{\mathrm{TP}}{\mathrm{TP}+\mathrm{FP}}$, and
$\mathrm{NPV}=\frac{\mathrm{TN}}{\mathrm{TN}+\mathrm{FN}}$.
Here, $\mathrm{Recall}$ measures the fraction of ground-truth vulnerabilities for which \sysname produces and validates an effective PoC. $\mathrm{PPV}$ measures the credibility of confirmations, i.e., the fraction of \sysname-confirmed cases that are truly exploitable. $\mathrm{NPV}$ measures the correctness of rejections, i.e., the fraction of \sysname-rejected cases that are indeed non-vulnerable. Together, these metrics capture both PoC reproduction effectiveness and the reliability of \sysname's validation decisions.

\begin{table}[t]
\centering
\caption{Evaluation of \sysname on benchmarks. We report TP/FP for Vul sets and TN/FN for Fix sets by manual verification. Execution Passed counts PoCs after the GRE-Engine stage, while \sysname Confirmed counts the final PoCs confirmed by \sysname.}

\footnotesize
\begin{tabular}{lcccccccc}
\toprule
\multirow{2}{*}{Dataset} & \multirow{2}{*}{Bugs}
& \multicolumn{2}{c}{Execution Passed}
& \multicolumn{2}{c}{\sysname Confirmed} \\
\cmidrule(lr){3-4}\cmidrule(lr){5-6}\cmidrule(lr){7-8}
& & DeepSeek & GPT
  & DeepSeek & GPT \\
\midrule
SmartBugs-Vul & 139 & 120/17 & 122/9 &  117/2 &  116/2  \\
FORGE-Vul     & 428 & 369/26 & 380/23 &  365/5  & 376/3  \\ \midrule
SmartBugs-Fix & 139  & -/64 & -/87 &  63/1   &  85/2    \\
FORGE-Fix     & 100  & -/57 & -/63 &   56/1  &   63/0    \\
\bottomrule
\end{tabular} 
\label{tab:rq1-total}

\end{table}

\begin{figure*}[!tbp]
  \centering
  \subfloat[Non-exploitable time manipulation.]{
    \includegraphics[width=0.45\linewidth]{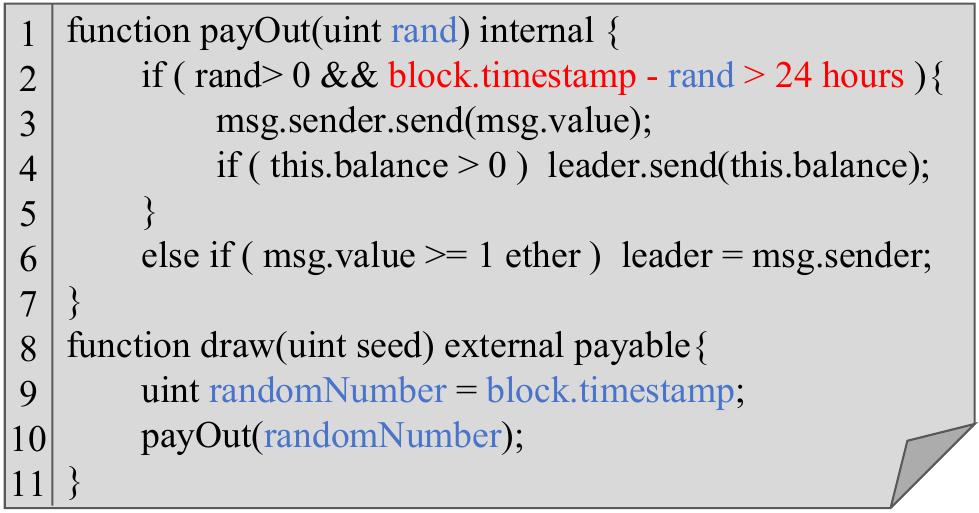}
    \label{fig:smartbugs-case}
  }\hfill
  \subfloat[False positive from ERC20 inheritance.]{
    \includegraphics[width=0.45\linewidth]{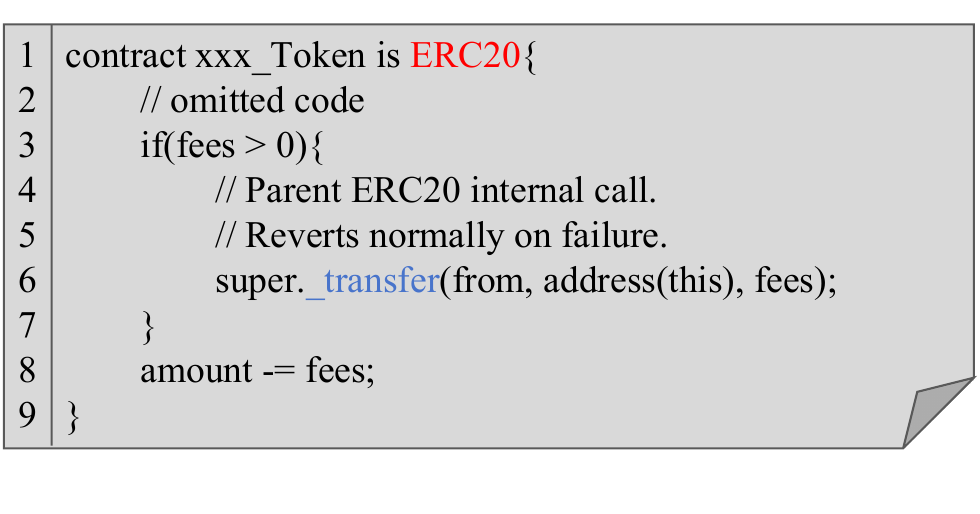}
    \label{fig:forge-case}
  }
  \caption{Representative inaccuracies in the benchmark ground truth.
  (a) \texttt{block.timestamp} is used, but the settlement condition is unreachable.
  (b) \texttt{super.\_transfer()} is an internal parent call that reverts on failure.}
  \label{fig:case-study}
\end{figure*}

\subsection{RQ1: Validation Accuracy on Vul/Fix Sets} \label{sec:rq1}

We evaluate whether \sysname can correctly validate reported vulnerabilities on vulnerable contracts while rejecting misleading reports on patched contracts in Table~\ref{tab:rq1-total}. All outcomes are independently adjudicated by two security experts and categorized as TP/FP for the Vul sets and TN/FN for the Fix sets.

Using the \sysname Confirmed column as the final outcome, \sysname achieves recalls of 84.17\% (117/139) and 85.28\% (365/428) on \textsc{SmartBugs-Vul} and \textsc{FORGE-Vul}, with corresponding PPVs of 98.32\% (117/119) and 98.65\% (365/370), respectively. On \textsc{SmartBugs-Fix} and \textsc{FORGE-Fix}, it attains NPVs of 99.28\% (138/139) and 99.00\% (99/100). Taken together, these results show that \sysname reliably validates real vulnerabilities while remaining conservative on patched code, thereby demonstrating its effectiveness in validating the correctness of static vulnerability reports.

\noindent \textbf{Effect of DV.}
The gap between Execution Passed and \sysname Confirmed highlights the benefit of DV. It increases PPV from 87.59\% (120/137) to 98.32\% on \textsc{SmartBugs-Vul} and from 93.42\% (369/395) to 98.65\% on \textsc{FORGE-Vul}. On the repaired sets, DV correctly rejects 63/64 and 56/57 executable but non-exploiting PoCs.

\noindent \textbf{Sensitivity to LLM backends and cost.}
Replacing DeepSeek-R1 with GPT-5-mini yields comparable recall, confirming 83.45\% (116/139) and 87.85\% (376/428) of cases on \textsc{SmartBugs-Vul} and \textsc{FORGE-Vul}, respectively. The average API cost is \$0.03 per validation with DeepSeek-R1 and \$0.04 with GPT-5-mini. The corresponding end-to-end latency is 52.14/60.47\,s and 13.84/19.68\,s on the two datasets, respectively.

Despite its strong overall performance, \sysname fails on a small subset of cases. Manual inspection attributes these failures to two factors: (i) the intrinsic difficulty of PoC synthesis and validation, and (ii) inaccuracies in the benchmark itself. We identify at least five such cases in \textsc{SmartBugs-Vul} and nine in \textsc{FORGE-Vul}.
Figure~\ref{fig:smartbugs-case} shows an example from \textsc{SmartBugs-Vul} labeled as \emph{Time Manipulation}. However, the claimed exploit path is infeasible: the settlement branch is unreachable because \texttt{block.timestamp} always equals \texttt{rand}. The reported vulnerability is therefore non-exploitable, and \sysname cannot generate a passing PoC.
Figure~\ref{fig:forge-case} shows a case from \textsc{FORGE-Vul} flagged as \emph{Unchecked Transfer}. However, the highlighted call is \texttt{super.\_transfer()}, an internal ERC-20 routine that reverts on failure rather than returning an unchecked Boolean value. DV therefore finds no execution evidence supporting the reported risk and rejects the case.
These examples suggest that \sysname can also help expose mislabeled or non-exploitable benchmark entries, providing useful signals for dataset curation.

\rqbox{1}{\sysname can effectively validate static vulnerability reports by transforming them into executable PoCs. 
Manual inspection further suggests that \sysname provides actionable signals for benchmark curation and re-annotation.}

\subsection{RQ2: Verification of Static-Analyzer Reports} \label{sec:rq2}

We run six static analyzers on \textsc{SmartBugs-Vul}, \textsc{FORGE-Vul} and \textsc{Latest-114}, validate the high-severity findings with \sysname, and manually audit sampled outcomes. As shown in Tables~\ref{tab:static-benchmarks}, \emph{Findings} denotes the number of high-severity reports produced by each analyzer, and \sysname denotes the subset confirmed by \sysname. PPV and NPV are determined through independent manual review by two experts.
Because manual validation is costly, we adjudicate a randomly sampled subset. We determine the minimum sample size using the standard finite-population proportion formula~\cite{lohr2021sampling}.
Specifically, we first compute the Cochran sample size for estimating a proportion under an infinite-population assumption with the standard settings (95\% confidence, margin of error $e=0.1$, conservative $p=0.5$):
\begin{equation}
n_0 = \frac{z_{\alpha/2}^{2}\, p(1-p)}{e^{2}},
\end{equation}
and then apply the finite population correction (FPC) to obtain the final sample size:
\begin{equation}
n = \frac{n_0}{1 + \frac{n_0 - 1}{N}},
\end{equation}
where $N$ is the number of findings produced by each static tool. 
We allocate the resulting sample budget between confirmed and rejected cases in proportion to their counts,
and then uniformly sample within each group to estimate $\mathrm{PPV}$ and $\mathrm{NPV}$.
When the computed $n$ is below 20, we round it up to 20 to ensure a minimum level of stability.

\begin{table}[!tbp]
\centering
\caption{Evaluation of static findings on benchmarks. Counts include findings reported by each static analyzer and those verified by \sysname. PPV/NPV are estimated via manual adjudication on random samples, with sample sizes computed using the finite-population proportion formula.}
\fontsize{6}{6}\selectfont
\begin{tabularx}{\linewidth}{XXcccc}
\toprule
Dataset & Analyzer & Findings & \sysname & $\mathrm{PPV}$ & $\mathrm{NPV}$ \\
\midrule
\multirow{6}{*}{\fontsize{5.5}{6}\selectfont\textsc{SmartBugs-Vul}}
& Slither & 91  & 44  & 16/22 (72.73\%) & 18/25 (72.00\%) \\
& Mythril & 156 & 87  & 23/33 (69.70\%) & 20/27 (74.07\%) \\
& CCC     & 287 & 143 & 27/36 (75.00\%)& 20/37 (54.05\%)\\
& GPTScan & 7   & 5   & 5/5 (100\%)  & 2/2 (100\%) \\
& DPO     & 61  & 32  & 15/20 (75.00\%) & 14/19 (73.68\%)\\
& FT      & 413 & 196 & 31/37 (83.78\%) & 33/42 (78.57\%) \\
\midrule
\multirow{6}{*}{\textsc{FORGE-Vul}}
& Slither & 621  & 47  & 17/20 (85.00\%) & 70/78 (89.74\%) \\
& Mythril & 4    & 0   & -     & 3/4 (75.00\%) \\
& CCC     & 0    & 0   & -     & - \\
& GPTScan & 56   & 10  & 9/10 (90.00\%) & 15/30 (50.00\%)\\
& DPO     & 1533 & 129 & 16/20 (80.00\%) & 83/84 (98.81\%)\\
& FT      & 363  & 96  & 14/20 (70.00\%) & 56/57 (98.25\%)\\
\midrule
\multirow{6}{*}{\textsc{Latest-114}}
& Slither & 133 & 62 & 20/26 (76.92\%) & 30/31 (96.77\%) \\
& Mythril & 9   & 5  & 5/5 (100\%)  & 4/4  (100.0\%)\\
& CCC     & 167 & 70 & 14/25 (56.00\%) & 36/37 (97.30\%)\\
& GPTScan & 0   & 0  & -     & - \\
& DPO     & 1   & 0  & -     & 1/1 (100\%) \\
& FT      & 235 & 99 & 25/29 (86.21\%)  & 39/40 (97.50\%) \\
\bottomrule
\end{tabularx}

\label{tab:static-benchmarks}
\end{table}

\begin{table*}[!tbp]
\centering
\caption{Total PPV/NPV of different static analyzers with their average report length.}
\small
\begin{tabular}{lcccccc}
\toprule
& CCC &Slither & Mythril  &  GPTScan& DPO & FT \\
\midrule

PPV & 41/61 (67.21\%) & 53/68 (77.94\%) & 28/38 (73.68\%) & 14/15 (93.33\%) & 31/40 (77.50\%) & 70/86 (81.40\%) \\
NPV & 56/74 (75.68\%)& 118/134 (88.06\%) & 27/35 (77.14\%) & 17/32 (53.13\%) & 98/104 (94.23\%) & 128/139 (92.09\%) \\
\midrule
Length &11.92& 36.42   & 54.26 & 101.54 &105.92 & 167.23 \\ 
\bottomrule
\end{tabular}
\label{tab:static-length}
\end{table*}

Across the three benchmark datasets, \sysname achieves 76.47\% (117/153), 80.00\% (56/70) and 75.29\% (64/85) PPV, along with NPV of 70.39\% (107/152), 89.72\% (227/253) and 97.35\% (110/113), respectively. These results demonstrate \sysnames ability to translate static-analysis findings into executable PoCs and to reliably validate or refute their exploitability. Manual inspection of successful PoCs and their corresponding reports shows that static analyzers typically highlight only a small set of functions, whereas working PoCs almost always require a richer call chain. In our effective PoCs, we consistently observe both classes of relationships introduced in Section~\ref{sec:bug-context-extraction}. For example, \emph{Mythril} may report “any sender can withdraw Ether from the contract account” and pinpoints \texttt{withdraw} as the vulnerable function. In practice, the exploit succeeds because a misconfigured modifier on \texttt{withdraw} delegates control to \texttt{\_withdraw}, the true sink, which is an instance of our \emph{structural links}. Moreover, the PoC must first call \texttt{transfer} to fund the attacker account before triggering the withdrawal logic, which is an instance of our \emph{semantic links}. This example illustrates that \textsc{BCE} can supply the additional code context required in real-world settings, enabling end-to-end PoCs that execute correctly.

In Table~\ref{tab:static-length}, we order the analyzers by their average report length and present the verification outcomes when each tool is paired with \sysname. We observe that traditional analyzers produce substantially shorter reports than LLM-based approaches. For example, CCC’s average report length is only 11.92, whereas FT’s is about 14$\times$ longer. During manual adjudication, we find that 70.00\% (14/20) of CCC’s false positives stem from the LLM misinterpreting the intent of overly terse descriptions. This suggests that report quality can affect \sysname’s downstream verifiability.

In addition, \sysname provides analyzers with a reproducible, execution-grounded verification signal.
For the two tools with relatively high NPV, namely DPO (94.23\%) and FT (92.09\%), we manually analyze the sources of their false positives. We find that most false positives stem from two factors.
(i) Overly conservative security warnings or audit recommendations that lack concrete execution paths. For example, these tools may report access-control issues even when the target function is already protected by an authorized modifier, or when the required checks have been performed in upstream callers.
(ii) Hallucinated reports that deviate substantially from the actual contract logic. Such hallucinations are often triggered by excessively long contract inputs or by limitations of the underlying models.

By contrast, GPTScan attains the highest PPV and the lowest NPV under our verification pipeline, suggesting higher precision and fewer spurious reports in this setting. 
We attribute this to GPTScan’s design, which combines LLM-based reasoning with additional static checks as a second validation gate. Overall, these results suggest that LLM-based analyzers can benefit from execution-grounded validation signals to corroborate their findings and reduce spurious alerts. \sysname provides a reproducible empirical signal for comparing how often reported findings are practically verifiable.

\rqbox{2}{\sysname can validate static-analysis findings by synthesizing executable PoCs and checking their exploitability. Moreover, by producing reproducible, execution-grounded evidence, \sysname helps surface limitations of existing analyzers, such as false positives that arise when findings are not corroborated by execution-grounded validation.}

\begin{table}[!tbp]
\centering
\caption{Comparison of PoC generation performance under different configurations on ground-truth datasets.}
\small
\begin{tabular}{lcc}
\toprule
Configuration & SmartBugs-Vul & FORGE-Vul \\
\midrule
\sysname & 137/139 (98.56\%) & 395/428 (92.29\%) \\
Agentic  & 132/139 (94.96\%) & 388/428 (90.65\%)  \\
Iterative &  72/139 (51.80\%) & 175/428 (40.89\%) \\
\midrule
w/o BCE & 121/139 (87.05\%) & 334/428 (78.04\%) \\
w/o Semantic Links & 103/139 (74.10\%) & 251/428 (58.84\%) \\
w/o Pre-Repair & 57/139 (41.01\%) & 149/428 (34.81\%) \\
w/o Post-Repair & 25/139 (17.99\%) & 61/428 (14.25\%) \\

\bottomrule
\end{tabular}
\label{tab:ablation-vul}
\end{table}

\subsection{RQ3: Performance of PoC Generation } \label{sec:ablation-study}

We evaluate PoC generation performance on SmartBugs-Vul and FORGE-Vul using DeepSeek-R1 (Table~\ref{tab:ablation-vul}). \sysname outperforms the two baselines by 3.60\%/1.64\% and 46.76\%/51.40\%.
We attribute the performance gap primarily to two factors relative to \emph{Iterative}. 
First, the baselines differ substantially in how they process the input code. 
The \emph{Iterative} baseline directly feeds the full contract code into the model. 
In contrast, \sysname uses BCE to expand the context along predefined inter-function relations, which more effectively identifies functions relevant to vulnerability triggering. 
This result suggests that structured bug-context extraction is more effective than using the full code context naively.
Second, \sysname benefits from the pre-execution sanitizer in the GRE-Engine. 
We observe that LLMs are often insensitive to compiler-version constraints and the APIs supported by each version. 
Such issues are difficult to resolve through iterative prompting alone, which highlights the necessity of our version-aware normalization and repair design. 
Although the rate of \sysname and \emph{Agentic} is similar, the \emph{Agentic} baseline must autonomously explore file paths and perform planning, resulting in input token costs and runtimes that are 4.96/5.35 and 14.81/15.53 times higher than those of \sysname, respectively.

We further conduct ablation studies on the major modules.

\noindent \textbf{BCE.}
Removing BCE reduces the generation rate by 11.51\% and 14.25\%. Compared to the configuration that keeps only call-graph expansion, \sysname improves the generation rate by 24.46\% and 33.45\%, highlighting the value of semantic expansion. The larger drop on \textsc{FORGE-Vul} further suggests that semantic links are more beneficial in realistic and complex projects.
We also examine the reduction in token cost. With BCE, the average input length on \textsc{FORGE-Vul} decreases from 33.67k to 16.71k tokens, corresponding to a 50.37\% reduction. On \textsc{SmartBugs-Vul}, it decreases from 5.29k to 4.72k tokens, a 10.78\% reduction.
Overall, these results indicate that BCE enables the model to focus more effectively on bug-relevant code while substantially reducing token cost.

\noindent \textbf{GRE-Engine.}
Ablating the pre-execution repair drops the rate to
41.01\% and
34.81\%, indicating that without targeted version mapping and import-path fixes, multi-round prompting alone is often insufficient to produce executable PoCs.
Ablating the post-execution repair—i.e., running only a single generation round with no feedback—reduces accuracy to
17.99\% and
14.25\%. 
These results show that both the pre-execution sanitizer and the post-execution debugger are crucial for successful PoC generation.

\rqbox{3}{\sysname outperforms the baselines in PoC generation. BCE and the GRE-Engine substantially increase the number of successfully generated PoCs by supplying precise context and repairing common generation and execution errors.}

%% file: 6-discussion.tex
\section{Threats to Validity} \label{sec:discussion}

\noindent \textbf{Soundness of LLM-Based Verification.} 
A potential threat to validity arises from our use of an LLM in the DV step to insert runtime checks, as errors in this instrumentation could affect exploitability judgments. However, the LLM does not directly determine exploitability; it only generates verification insertions, whereas the final judgment is based on concrete runtime evidence. DV is therefore execution-grounded rather than an LLM-as-a-judge mechanism.
As shown in Table~\ref{tab:rq1-total}, expert adjudication yields PPVs of 98.32\% on \textsc{SmartBugs-Vul} and 98.65\% on expert audit reports \textsc{FORGE-Vul}, indicating that confirmations are highly reliable in practice. This finding also indirectly supports the reliability of the LLM-assisted verification step, which is substantially more constrained than end-to-end PoC generation. Nevertheless, residual risk remains in edge cases where hallucinations produce incomplete, imprecise, or semantically misaligned checks.

\noindent \textbf{Non-faithful PoC Generation.}
An LLM may generate a passing PoC by altering the target semantics rather than faithfully exploiting the original vulnerability. Such ``cheating'' behavior may include redefining key functions, bypassing the intended execution path, or introducing auxiliary logic that makes validation succeed for the wrong reason. We mitigate this risk through deterministic pre-execution sanitization, but subtle semantic deviations may still escape our checks. Thus, while our design substantially reduces the risk of non-faithful validation, it cannot rule it out entirely.

%% file: 7-related-works.tex
\section{Related Work}

\textbf{Formal verification.}
Formal methods encode smart-contract safety properties and use SMT solving, bounded model checking, or symbolic execution to prove or refute them. Some works~\cite{so2021smartest,mueller2018smashing,mossberg2019manticore,krupp2018teether,nikolic2018maian,frank2020ethbmc,permenev2020verx}
systematically explore transaction sequences under user-defined or built-in properties, producing proofs or counterexample traces.
More recent systems such as SmartInv~\cite{wang2024smartinv} and PropertyGPT~\cite{liu2024propertygpt} use multimodal prompting and retrieval-augmented generation to synthesize candidate invariants, reducing manual specification effort.
However, these approaches remain fundamentally proof-centric, delivering formal proofs or counterexample traces rather than executable PoC tests. As such, they necessitate complementary dynamic validation in realistic environments to confirm that identified violations translate to real exploitability.

\textbf{Fuzzing and stateful exploration.}
Fuzzing-based techniques dynamically explore smart-contract behavior and have proven effective at uncovering exploitable traces. Early systems such as ContractFuzzer~\cite{jiang2018contractfuzzer}, Echidna~\cite{grieco2020echidna}, and EthPloit~\cite{zhang2020ethploit} pioneered ABI-/grammar-based and property-based fuzzing for Ethereum contracts. More recent work, including ItyFuzz~\cite{shou2023ityfuzz}, VERITE~\cite{kong2025verite}, Midas~\cite{ye2024midas}, and  Ethploit~\cite{zhang2020ethploit}, targets deep stateful behavior and economically exploitable traces using snapshot-based execution, guided seed selection, and profit-aware feedback to synthesize concrete exploit transactions. These dynamic systems rely on runtime signals as oracles and are highly discovery-oriented. However, they are driven by program code or on-chain state rather than generic static audit reports, and their oracles are often tailored to specific bug classes, limiting their applicability as general verification mechanisms for arbitrary static findings.

\textbf{LLM-assisted exploit and test generation.}
Large language models have been widely applied to code synthesis, including the generation of validation tests and exploit scripts. Some exploit-generation approaches require specialized inputs, such as transaction histories, exploit templates, or counterexamples, and are typically designed for specific exploitation settings rather than general static-analysis findings~\cite{zhang2025automated,sun2025pocshift,wei2025veriexploit,wu2024advscanner,so2021smartest,fang2023isyn,eshghie2024oracle,su2026transactions}. More recent studies~\cite{xiao2025prompt,gervais2025aiagentsmartcontract} investigate the use of LLMs to discover vulnerabilities and synthesize exploits directly from source code. PoCo~\cite{andersson2025poco} adopts an agentic loop to generate PoCs from auditor-written vulnerability annotations and iteratively refines them using compilation and test feedback.
\sysname differs from these approaches in that it targets the dynamic validation of heterogeneous static-analysis findings. Existing pipelines either depend on stronger task-specific inputs or emphasize PoC/exploit synthesis rather than exploitability validation from raw findings. They also generally lack a generic automated runtime oracle for determining whether exploitation has succeeded across diverse bug classes without manually specified success predicates. As a result, the validation of heterogeneous static-analysis outputs remains an open practical challenge.

%% file: 8-conclusion.tex
\section{Conclusion}
In this work, we propose a dynamic validation approach for smart-contract static findings by generating executable, validated PoCs. We implement this approach in a tool called \sysname, which addresses noisy inputs, lack of execution grounding, and automated runtime verification through its BCE, GRE-Engine, and DV modules. Across benchmarks and real-world contracts, \sysname converts textual reports into validated PoCs, achieving 84.17\% and 85.28\% recall rate on \textsc{SmartBugs-Vul} and \textsc{FORGE-Vul}, respectively, and confirming 64 vulnerabilities on the latest Ethereum corpus at an average cost of \$0.03 per finding. Looking ahead, we release a curated set of validated PoCs to better guide LLMs in learning and generating verification code.

%% file: 9-data.tex
\section{Ethics Considerations.}
All experiments were conducted in isolated environments on forked mainnet snapshots. We did not interact with live contracts, hold real assets, or submit on-chain transactions. For the Etherscan verified-source corpus, we manually reviewed all findings and found no evidence of active deployments holding exploitable tokens at the time of our submission. We reported potential issues to deployers via Etherscan/Blockscan Chat and withheld exploit details, PoC code, and contract identifiers unless explicit consent is obtained.